\def\year{2022}\relax
\documentclass[letterpaper]{article} 
\usepackage{aaai22}  
\usepackage{times}  
\usepackage{helvet}  
\usepackage{courier}  
\usepackage[hyphens]{url}  
\usepackage{graphicx} 
\usepackage{natbib}  
\usepackage{caption} 
\DeclareCaptionStyle{ruled}{labelfont=normalfont,labelsep=colon,strut=off} 
\usepackage{algorithm}
\usepackage{algorithmic}

\usepackage{multirow}
\usepackage{comment}

\usepackage{xcolor}
\newcommand{\answerYes}[1]{\textcolor{blue}{#1}} 
\newcommand{\answerNo}[1]{\textcolor{teal}{#1}} 
\newcommand{\answerNA}[1]{\textcolor{gray}{#1}}

\usepackage{newfloat}
\usepackage{listings}
\floatstyle{ruled}
\newfloat{listing}{tb}{lst}{}
\floatname{listing}{Listing}
\title{Detecting Cultural Differences in News Video Thumbnails \\ via Computational Aesthetics}

\author{
    Marvin Limpijankit \textsuperscript{\rm 1}, John R. Kender \textsuperscript{\rm 1}
}
\affiliations{
    \textsuperscript{\rm 1}Columbia University\\

    New York, NY 10027 \\
    ml4431@columbia.edu, jrk@cs.columbia.edu
}

\usepackage{bibentry}

\begin{document}

\maketitle

\begin{abstract}
We propose a two-step approach for detecting differences in the style of images across sources of differing cultural affinity, where images are first clustered into finer visual themes based on content before their aesthetic features are compared. 
We test this approach on 2,400 YouTube video thumbnails taken equally from two U.S. and two Chinese YouTube channels, and relating equally to COVID-19 and the Ukraine conflict. 
Our results suggest that while Chinese thumbnails are less formal and more candid, U.S. channels tend to use more deliberate, proper photographs as thumbnails.  
In particular, U.S. thumbnails are less colorful, more saturated, darker, more finely detailed, less symmetric, sparser, less varied, and more up close and personal than Chinese thumbnails.
We suggest that most of these differences reflect cultural preferences, and that our methods and observations can serve as a baseline against which suspected visual propaganda can be computed and compared.
\end{abstract}

\section{Introduction}


Visual imagery
is an extremely important channel for international news coverage, as it connects the audience to events that may otherwise be removed from direct experience, 
and serves as a powerful tool for conveying emotion~\cite{smith2013public}.
For threatening events, such as global warming, wars, or pandemics, images have also been shown to be significant drivers of public engagement, “providing viewers with tangible and emotion-evoking examples that act as visual proof”~\cite{joffe2008power, martikainen2021newspaper}.
As such, highly visual mediums like YouTube
play a significant role in modern news consumption~\cite{burgess2018youtube, xie2011visual}.

In this paper, we
apply computational approaches to quantify 
differences in the use of visual imagery between news sources from varying regional backgrounds.
More specifically, our work lies at the interplay of sociology and computer science, adopting techniques from computer vision to extract visual features from YouTube thumbnails at scale, before using methods such as dimensionality reduction to accentuate cross-cultural differences between U.S. and Chinese channels.  
We focus on two international events, the COVID-19 pandemic, and the war in Ukraine.

Since statistical image properties are highly dependent on the content of the image~\cite{kao2017deep, machajdik2010affective},
our proposed framework first
clusters images about a given news story by ``visual themes'' (i.e. different aspects or sub-events), 
We then analyze the aesthetic differences between channels.
Lastly, we consider the views, likes, and comments for each video, in order to capture
potential cross-cultural differences in how thumbnail properties may influence video performance.

Our contributions in this paper are:
\begin{enumerate}
  \item A framework to compare image properties in YouTube thumbnails that accounts for visual thematic content.
  \item An algorithmic extraction of 21 visual aesthetics of a dataset of 2,400 thumbnail images, taken from two U.S. and two Chinese channels, covering two separate international events, COVID-19 and the Ukraine war. 
  \item An analysis of the culture-specific differences in the use of these aesthetics, and their relation to viewership statistics and potential disinformation.
\end{enumerate}

\section{Related Work}

\subsection{Computational Aesthetics} 

In automatic image aesthetic assessment, also known as computational aesthetics, a model is trained to quantify the beauty of an image through some aesthetic score.
Often, these models assess an image by measuring the degree to which it adheres to fundamental
but abstract 
principles of photography, such as 
balance, rhythm, harmony, unity, tone, etc.~\cite{anwar2021survey}.

There are two main approaches to image aesthetic assessment: hand-crafted methods, and deep learning methods. 

\subsubsection{Hand-crafted methods.}

These rely on extracting pre-defined low-level features from images and fitting a machine learning model, such as SVM, to do the prediction. 
Examples of features that have been extensively explored previously include sharpness, colorfulness, contrast, and texture~\cite{lo2013intelligent, redi2015beauty, aydin2014automated}.

\subsubsection{Deep learning methods.}

These leverage neural networks to learn low and high-level image features without (or with relatively little) human instruction. 
These are trained on large image datasets such as Photo.net~\cite{datta2006studying}, 
AVA~\cite{murray2012ava},
and CUHK-PQ~\cite{luo2011content},
and the specific architecture can vary. 
Doshi et al.~\cite{doshi2020image} fine-tuned two pre-trained models (AlexNet and VGG) to classify images into high and low aesthetic categories. 
Kao et al.~\cite{kao2017deep} investigated using a Multi-Task Convolutional Neutral network (MTCNN) to jointly perform aesthetic assessment and semantic recognition. 

\subsubsection{Applications.}

In addition to predicting image aesthetics such as quality, beauty, and liking~\cite{bartho2023predicting, peng2018feast},
there is evidence to suggest that computational aesthetics can serve as a valuable starting point for a variety of downstream tasks. 

For instance, one experiment demonstrated the ability to infer the personality traits (``Big Five Inventory'') of Flickr users based on the color, composition, texture, and face features of images that the user tagged as ‘favorites’ on the platform~\cite{segalin2016pictures}. 

Likewise, Machajdik et al.~\cite{machajdik2010affective} achieved state-of-the-art results on image emotion classification using a similar set of features, including color, texture, composition, and content, inspired by psychology and art theory. 
However, results were not uniform:
content features such as the number of faces and relative size of the biggest face were good discriminators between fear/disgust vs. amusement in the International Affective Picture System
dataset, whereas color features was more predictive
in an artistic photography dataset collected from the DeviantArt website.

These studies reflect the ability of image aesthetics to emphasize certain feelings, but warn of the importance of considering the content and source.

\subsection{Image Analysis in News Media} 

Cross-cultural differences in news portrayals have been studied extensively from both a 
sociological (manual)
and computational (automatic) perspective. 

\subsubsection{Manual.} 

One investigation analyzed images of the COVID-19 pandemic appearing in Finnish newspapers, focusing on the social representations of individuals from varying age groups~\cite{martikainen2021newspaper}. 
The authors noted that the angle of the photo often contributed to visual rhetoric, for instance, images of children studying at home emphasized a high angle, placing the spectator in the authoritative position overseeing the child whereas those of young adults tended to be taken at eye-level. 

Similarly, in a study on how the media covered Pope Francis’ visit to Cuba, Thomson et al. discovered that photo angles in U.S. sources framed the Pope at higher positions than Cuban politicians whereas local sources used a more equal leveling~\cite{thomson2018politicians}. 
Another contrast was the content of the images, with Cuban media focusing almost exclusively on the Pope, whereas U.S. media more often published photographs of Cuban people.
~\cite{fahmy2010contrasting, rafiee2023framing}.

\subsubsection{Computational.} 

\begin{figure}[ht]
\begin{center}
\includegraphics[height=4cm, width=8cm]{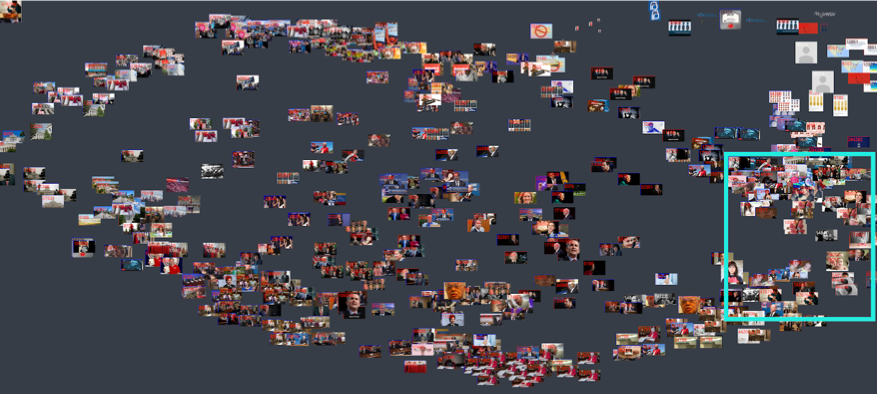}
\caption{An example of Photo-Space visualization~\cite{bhargava2020mapping}, where physical placement captures thematic relations, and image borders are color-coded by political affinity.}
\label{fig:photo-space}
\end{center}
\end{figure}

Although these frequently involve analyzing text data to compare sentiments and topics~\cite{wallbing2021computational, chen2023leveraging}, the visual medium of news media has also been explored. 
Bhargava et al.~\cite{bhargava2020mapping} proposed Photo Spaces, a visualization tool to describe visual narratives for an event, based on image embeddings produced by a pre-trained ResNet-50 model. 
The spatial positioning of images in the photo-space is determined by reducing these embeddings into two dimensions, via
the Uniform Manifold Approximation and Projection technique, from which clusters corresponding to visual themes can be extracted. 
The authors document an example of the tool (See Figure~\ref{fig:photo-space}) on abortion rights as a topic, adding the political affinity of sources (right- vs. left-leaning)
through the color of each image’s border.

Similarly, Zhang et al.~\cite{zhang2022image} evaluated the effectiveness of bag-of-visual-words, self-supervised learning, and transfer learning to cluster unlabeled image data. 
They conclude that transfer learning using pre-trained models to obtain embeddings outperforms the other methods, though the choice of the pre-trained model
significantly affected clustering results. 

Following these trends, 
in our investigation we adopt image clustering as a pre-processing step in our pipeline, 
as a way of controlling for visual context.

\subsection{Video Thumbnails}

Thumbnails are small image representations of full-length videos.
Thumbnails ``tell the users what the video is about (i.e., be informative), and grab potential viewers' attention (i.e., be visually appealing)''~\cite{koh2022exploration}. 

Kim et al.~\cite{kim2023towards} evaluated the design of thumbnails for data-driven news articles, concluding similarly that for thumbnails involving informative graphics, users prefer ones that are interpretable and attractive. 
For branded content, a thumbnail’s success (measured in view-through) can also be attributed to visual features, such as colorfulness and brightness~\cite{koh2022exploration}. 

As such, 
our work pursues the observations that
thumbnails may serve as an interesting visual medium for cross-cultural news analysis, reflecting both the content the media wishes to present (information) as well as their attempts to appeal to specific cultural perspectives (attraction).

\section{Methods}

\subsection{Data Collection}

\subsubsection{Choice of channel.}

To evaluate cross-cultural differences, we selected two U.S. and two Chinese YouTube channels. 
The U.S. sources are ABC and CBS, with 15.6M and 5.5M subscribers respectively, and the Chinese sources are CGTN (Chinese Global Television Network) and New China TV, with 3.0M and 1.4M subscribers respectively. 

\subsubsection{Choice of event.}

Inspired by previous work~\cite{martikainen2021newspaper},  
we selected the COVID-19 pandemic as one of the international news events to examine.
The complexity of the event gives rise to a
diverse set of event themes---medical, economic, political---which may transfer to detectable visual themes.
For similar reasons, we also selected the war in Ukraine, a more recent event 
with different event themes---military conflict, the humanitarian crisis, and geopolitics. 

\subsubsection{Thumbnail curation.}

The YouTube API
was used to scrape the video metadata using a search query (``covid 19'', ``ukraine war''), the channel\_ids of the above, and a published\_after date. 
For each event and each channel, we collected the top 300 videos, sorted by YouTube's default ``relevance'' parameter.
Iterating through the results, image content is then scraped using a simple Python script, and pre-processed to remove any vertical and horizontal black bars using the Pillow package. 
Each of the resulting 2,400 thumbnails were recorded, associated with an image\_id, channel, view count, like count, and comment data.

\subsection{Visual Theme Identification}

Prior to any comparative analysis, in order to control for image content, the thumbnails are first divided into subgroups based on visual themes. 

\subsubsection{Thumbnail clustering.}

\begin{figure}[ht]
\begin{center}
\includegraphics[height=3.9cm]{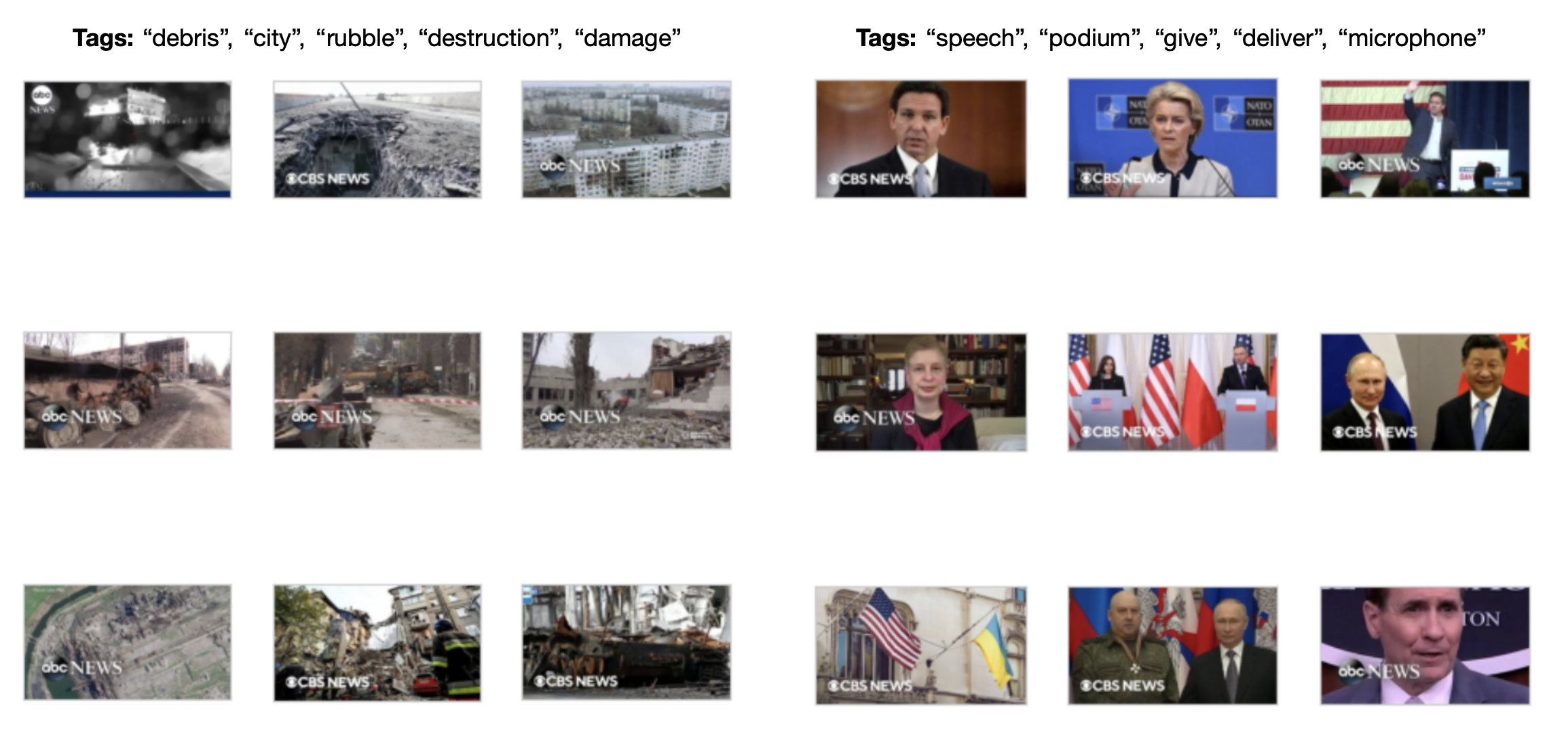}
\caption{Example visual themes from the Ukraine war.}
\label{fig:themes}
\end{center}
\end{figure}

After some experimentation, our best results were obtained by using
Concept~\cite{grootendorst2024concept}, a CLIP and BERTopic-based open source tool that performs topic modeling on images, by first encoding images using CLIP, and then identifying clusters on top of those embeddings. An example of the results produced using this method is shown in Figure~\ref{fig:themes}. 

\begin{table}[ht]
\resizebox{\columnwidth}{!}{%
\begin{tabular}{|c|c|l|l|l|}
\hline
\multirow{2}{*}{Event}       & \multirow{2}{*}{Theme} & \multicolumn{3}{c|}{Tagging Method}                                                                                                                                                                                                                                        \\ \cline{3-5} 
                             &                          & \multicolumn{1}{c|}{\begin{tabular}[c]{@{}c@{}}Approach 1\\ (top 5)\end{tabular}} & \multicolumn{1}{c|}{\begin{tabular}[c]{@{}c@{}}Approach 2\\ (top 5 adjusted)\end{tabular}} & \multicolumn{1}{c|}{\begin{tabular}[c]{@{}c@{}}Approach 3\\ (tf-idf)\end{tabular}}        \\ \hline
\multirow{3}{*}{COVID}       & 0                        & \begin{tabular}[c]{@{}l@{}}woman, news, person, \\ stand, man\end{tabular}        & \begin{tabular}[c]{@{}l@{}}interview, sit, table,\\ blue, speech\end{tabular}              & \begin{tabular}[c]{@{}l@{}}mall, tie, interview,\\ sunglasses, give\end{tabular}          \\ \cline{2-5} 
                             & 1                        & \begin{tabular}[c]{@{}l@{}}news, wear, tie,\\ business suit, man\end{tabular}     & \begin{tabular}[c]{@{}l@{}}business suit, tie, suit,\\ interview, microphone\end{tabular}   & \begin{tabular}[c]{@{}l@{}}bookshelf, office, interview,\\ black, tie\end{tabular}        \\ \cline{2-5} 
                             & 2                        & \begin{tabular}[c]{@{}l@{}}stand, mask, man,\\ wear, person\end{tabular}          & \begin{tabular}[c]{@{}l@{}}mask, equipment, syringe,\\ hand, liquid\end{tabular}           & \begin{tabular}[c]{@{}l@{}}worker, tablet, syringe,\\ hospital room, garment\end{tabular} \\ \hline
\multirow{4}{*}{Ukraine war} & 0                        & \begin{tabular}[c]{@{}l@{}}business suit, speech, stand,\\ tie, man\end{tabular}  & \begin{tabular}[c]{@{}l@{}}speech, podium, give,\\ deliver, microphone\end{tabular}        & \begin{tabular}[c]{@{}l@{}}give, podium, business suit,\\ speech, tie\end{tabular}        \\ \cline{2-5} 
                             & 1                        & \begin{tabular}[c]{@{}l@{}}business suit, wear, tie,\\ stand, man\end{tabular}    & \begin{tabular}[c]{@{}l@{}}business suit, news, interview,\\ woman, suit\end{tabular}      & \begin{tabular}[c]{@{}l@{}}shake, suit, interview,\\ business suit, tie\end{tabular}      \\ \cline{2-5} 
                             & 2                        & \begin{tabular}[c]{@{}l@{}}man, rubble, debris,\\ city, building\end{tabular}     & \begin{tabular}[c]{@{}l@{}}debris, city, rubble,\\ destruction, damage\end{tabular}        & \begin{tabular}[c]{@{}l@{}}explosion, flame, damage,\\ garbage, rubble\end{tabular}       \\ \cline{2-5} 
                             & 3                        & \begin{tabular}[c]{@{}l@{}}stand, army, camouflage,\\ person, man\end{tabular}    & \begin{tabular}[c]{@{}l@{}}camouflage, army, soldier,\\ gun, equipment\end{tabular}        & \begin{tabular}[c]{@{}l@{}}weapon, load, gun,\\ soldier, rifle\end{tabular}               \\ \hline
\end{tabular}
}
\caption{Methods for theme tagging; we chose the second.}
\label{tab:tagging}
\end{table}

\subsubsection{Theme tagging.}

Thumbnail clusters themselves carry no easily interpretable meaning, so to understand the visual themes being extracted we associated each cluster with five textual tags. 
We leveraged Tag2Text~\cite{huang2023tag2text}, a model capable of recognizing 3,429 commonly human-used categories, to generate tags for every image.
Then, on a cluster level, we aggregated these tags to generate five tags that are most representative of the cluster's theme.

For this, we examined three approaches:
\begin{enumerate}
    \item Use the top five most commonly occurring tags. 
    \item Use the same approach as the first but with the added requirement that these tags appear less than some $k$ times across all images, including other clusters.
    \item Compute a tf-idf (term frequency-inverse document frequency) score for each tag with respect to a cluster using Equation~\ref{eq:tagweight}, and use the five tags with the largest weights.
\end{enumerate}
\begin{equation}
    w_{i, j} = tf_{i, j} \log \left ( \frac{N}{df_{i}} \right )
\label{eq:tagweight}
\end{equation}
 where $w_{i,j}$ is the tf-idf weight of tag $i$ in cluster $j$, $tf_{i, j}$ is the frequency of the tag in the cluster, $N$ is the number of clusters, and $df_{i}$ is the number of clusters containing the tag. 

\begin{figure}[ht]
\begin{center}
\includegraphics[height=6cm]{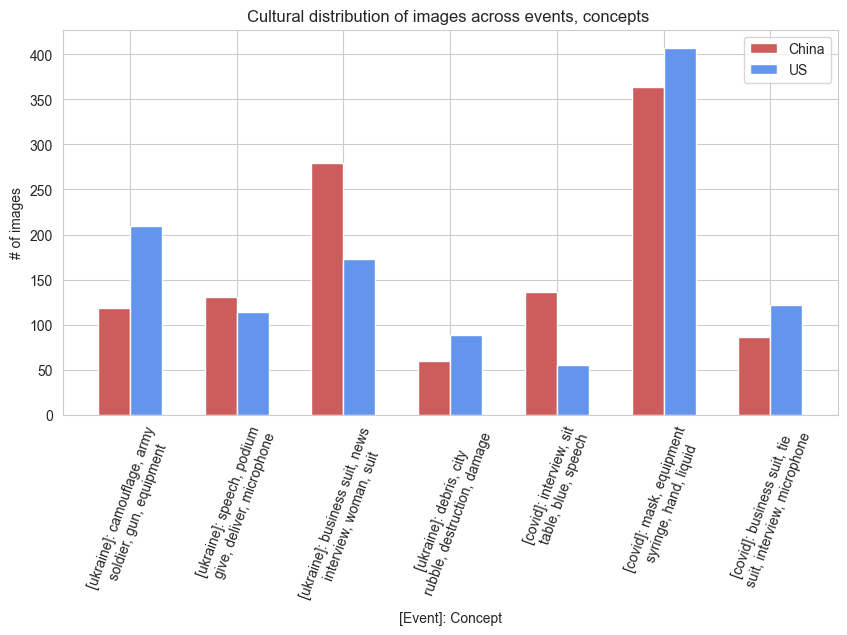}
\caption{Ukraine and COVID themes, China vs. U.S.}
\label{fig:cultural_distributions}
\end{center}
\end{figure}

The second and third approaches were introduced to filter out certain tags, such as {\tt man}, that appear in almost every thumbnail and thus are not discriminative between clusters. 
The results from these tagging methods respectively are summarized in Table~\ref{tab:tagging}.
We selected the second approach, which are charted in Figure~\ref{fig:cultural_distributions}.

Those four Ukraine themes, roughly ``speech'', ``interview'', ``rubble'', and ``army'', were distributed in a 2:4:1:3 ratio overall,
heavily favoring talk over destruction, and with Chinese thumbnails even more so.

Those three COVID themes, roughly ``speech'', ``interview'', and ``equipment'', were distributed in a 1:1:4 ratio overall, 
heavily favoring ``equipment'' by both cultures.
But Chinese thumbnails favored ``speech'' over ``interview''; U.S. thumbnails were the reverse.

\subsection{Image Feature Extraction}

Image features are extracted using a hand-crafted approach, where various pre-defined statistical measures are computed for each image.
An exhaustive reference list of the 21 features, in eight feature groups, 
is displayed in Table~\ref{tab:extracted}.
The majority of our feature set is adopted from Bartho et al.~\cite{bartho2023predicting}. 
A brief 
justification of each feature is included here, but the reader is encouraged to consult the references for the computational details and tuning parameters of each algorithm.

\begin{table*}[ht]
\resizebox{510.74pt}{!}{%
\begin{tabular}{|llcll|}
\hline
Group                            & Literature Reference(s)                                                                                                                                                                                                                                                    & \multicolumn{1}{l}{Dimensions}   & Feature                    & Description                                                                                                                            \\ \hline
\multirow{5}{*}{Color}           & \multirow{5}{*}{\begin{tabular}[c]{@{}l@{}}
\cite{datta2006studying}, \cite{li2009aesthetic},\\ \cite{mallon2014beauty},\\ 
\cite{schifanella2015image},\\ 
\cite{thieleking2020art},\\  
\cite{iigaya2021aesthetic}, \cite{geller2022statistical}
\end{tabular}} & \multirow{5}{*}{5}               & \textit{Hue}               & mean of hue channel in HSV space                                                                                                       \\
                                 &                                                                                                                                                                                                                                                                            &                                  & \textit{Saturation}        & mean of saturation channel in HSV space                                                                                                \\
                                 &                                                                                                                                                                                                                                                                            &                                  & \textit{Lab(a)}            & mean of a* channel in L*a*b* space                                                                                                         \\
                                 &                                                                                                                                                                                                                                                                            &                                  & \textit{Lab(b)}            & mean of b* channel in L*a*b* space                                                                                                         \\
                                 &                                                                                                                                                                                                                                                                            &                                  & \textit{Color Entropy}     & Shannon entropy of hue channel in HSV space                                                                                            \\ \hline
\multirow{2}{*}{Dimension} & \multirow{2}{*}{\begin{tabular}[c]{@{}l@{}}
\cite{datta2006studying},
\cite{iigaya2021aesthetic}\\
\cite{mallon2014beauty}
\end{tabular}}                                                                                                                                & \multirow{2}{*}{2}               & \textit{Aspect Ratio}      & image width divided by image height (pixels)                                                                                           \\
                                 &                                                                                                                                                                                                                                                                            &                                  & \textit{Image Size}        & image width + image height (pixels)                                                                                                    \\ \hline
\multirow{3}{*}{Lightness}           & \multirow{3}{*}{\begin{tabular}[c]{@{}l@{}}
\cite{schifanella2015image},\\
\cite{sidhu2018prediction}, 
\cite{mather2018visual},\\
\cite{iigaya2021aesthetic},
\cite{peli1990contrast}
\end{tabular}}                                                                           & \multirow{3}{*}{3}               & \textit{Contrast}          & standard deviation of L* channel in L*a*b* space                                                                                           \\
                                 &                                                                                                                                                                                                                                                                            &                                  & \textit{Luminance}         & mean of L* channel in L*a*b* space                                                                                                         \\
                                 &                                                                                                                                                                                                                                                                            &                                  & \textit{Luminance Entropy} & Shannon entropy of L* channel in L*a*b* space                                                                                              \\ \hline
\multirow{3}{*}{HOG}             & \multirow{3}{*}{\begin{tabular}[c]{@{}l@{}}
\cite{bosch2007representing},\\
\cite{braun2013statistical}
\end{tabular}}                                                                                                                                                          & \multirow{3}{*}{3}               & \textit{Self-Similarity}   & similarity of HOG features                                                                                                             \\
                                 &                                                                                                                                                                                                                                                                            &                                  & \textit{Complexity}        & mean gradient strength                                                                                                                 \\
                                 &                                                                                                                                                                                                                                                                            &                                  & \textit{Anisotropy}        & standard deviation of HOG features                                                                                                     \\ \hline
\multirow{2}{*}{Fourier}         & \multirow{2}{*}{\cite{redies2008fractal}}                                                                                                                                                                                                                               & \multirow{2}{*}{2}               & \textit{Fourier Slope}     & \begin{tabular}[c]{@{}l@{}}slope of line of best fit on the log-log plot of the \\ Fourier power spectrum\end{tabular}                 \\
                                 &                                                                                                                                                                                                                                                                            &                                  & \textit{Fourier Sigma}     & \begin{tabular}[c]{@{}l@{}}RMSE of line of best fit on the log-log plot of the \\ Fourier power spectrum\end{tabular}                  \\ \hline
\multirow{2}{*}{Symmetry}        & \multirow{2}{*}{\begin{tabular}[c]{@{}l@{}}
\cite{brachmann2016using}\\
\cite{bertamini2019study}
\end{tabular}}                                                                                                                                           & \multirow{2}{*}{2}               & \textit{Symmetry-LR}       & \begin{tabular}[c]{@{}l@{}}left-right symmetry based on first layer activations \\ on pre-trained AlexNet\end{tabular}                 \\
                                 &                                                                                                                                                                                                                                                                            &                                  & \textit{Symmetry-UD}       & \begin{tabular}[c]{@{}l@{}}up-down symmetry based on first layer activations \\ on pre-trained AlexNet\end{tabular}                    \\ \hline
\multirow{2}{*}{CNN}             & \multirow{2}{*}{\cite{brachmann2017using}}                                                                                                                                                                                                                            & \multirow{2}{*}{2}               & \textit{Sparseness}        & \begin{tabular}[c]{@{}l@{}}median variance of each max-pooled response map \\ from the first layer of pre-trained AlexNet\end{tabular} \\
                                 &                                                                                                                                                                                                                                                                            &                                  & \textit{Variability}       & \begin{tabular}[c]{@{}l@{}}variance over all max-pooled response maps\\ from the first layer of pre-trained AlexNet\end{tabular}       \\ \hline
\multirow{2}{*}{Other}           & \multirow{2}{*}{\begin{tabular}[c]{@{}l@{}}
\cite{savardi2018shot}, \\
\cite{zhang2022image}
\end{tabular}}                                                                                                                                                      & \multirow{2}{*}{\textgreater{}1} & \textit{Shot Scale}        & close, medium, or long shot based on fine-tuned CNN                                                                                    \\
                                 &                                                                                                                                                                                                                                                                            &                                  & \textit{Bag of Objects}    & frequency of objects detected by YOLOv5                                                                                                \\ \hline
\end{tabular}
}
\caption[]{\tabular[t]{@{}l@{}} Eight groups of
 visual features extractable from thumbnails. The first seven are scalar; the last is discrete.\\ \  \endtabular}
\label{tab:extracted}
\end{table*}

\subsubsection{Color.}

\textit{Hue, Saturation, Lab(a), Lab(b), Color Entropy}.
The use of colors can affect the perceived mood of an image.
Although there are multiple ways to represent colors numerically,
the HSV space (for hue, saturation, and value), and the L*a*b* space (for lightness, green-red, and blue-yellow) both
have been shown to be more aligned with the human perception of color~\cite{fadaei2021comparision}.
Color-entropy captures the visual percept of the ``colorfulness'' of an image.

\subsubsection{Dimension.}

\textit{Aspect Ratio, Image Size}. 
Although thumbnails are reshaped into pre-defined dimensions by YouTube,
videos vs. shorts are presented slightly differently.
Channels may also choose to include border padding within the thumbnail image itself, adding variation to dimension features since this is cropped in pre-processing.

\subsubsection{Lightness.}

\textit{Contrast, Luminance, Luminance Entropy}.
To complement color, several features for lightness (i.e. brightness) are also considered.
For instance, the effect of a very bright vibrant red such as Carmine is not be equivalent to that of a darker red such as Burgundy.

\subsubsection{HOG.}

\textit{Self-Similarity, Complexity, Anisotropy}.
A Histogram of Oriented Gradients (HOG) is a feature descriptor that extracts texture information, originally introduced for object recognition.
The method captures statistics about the distribution of the angles of color edges.
It first establishes a histogram whose bins span a quantized set of orientations, and then accumulates within each bin the L*a*b* gradient magnitudes that occur at that angle.  Each histogram is local to 
a single cell of a mesh-like partition of an image, and it therefore captures the edge activity within a local region.
These orientation statistics are then used to determine approximations to three more subjectively meaningful visual percepts.

\textit{Self-similarity} measures the overall likeness of the HOG features of a localized region of an image compared to its neighboring regions.
It is a useful feature for separating visually noisy objects from plain or even complicated backgrounds.

\textit{Complexity} is calculated as the average color gradient strength throughout an image.  
An image with many sharp color edges, and therefore very strong gradients, has high complexity; cartoons have low complexity.

\textit{Anisotropy} measures the relative strengths of gradients at different orientations, calculated as the standard deviation of HOG histogram bin magnitudes.
High values would suggest strongly differing strengths at different orientations, and are typical of human-created objects with parallel and perpendicular edges.

\subsubsection{Fourier.}

\textit{Fourier Slope, Fourier Sigma}.
The Fourier Transform
maps an image to the frequency domain, decomposing the image into a sum of two-dimensional sine and cosine components.
Low-frequency components capture less variant parts of the image, whereas high-frequency components capture sharp details.  
The structure of the image as a whole can be represented by charting how rapidly the logarithm of the
power (magnitude squared) of the Fourier transformation
decreases 
from low-frequency components to high-frequency components.

\textit{Fourier slope} is calculated as the slope of the line of best fit, whereas the \textit{Fourier sigma} is calculated as the root mean squared error of that line,
An image with larger (less negative) slope suggests a stronger presence of details, often natural, whereas smaller values suggest an overall smoother image, often man-made.

\begin{figure*}[ht]
\begin{center}
\includegraphics[height=5.5cm]{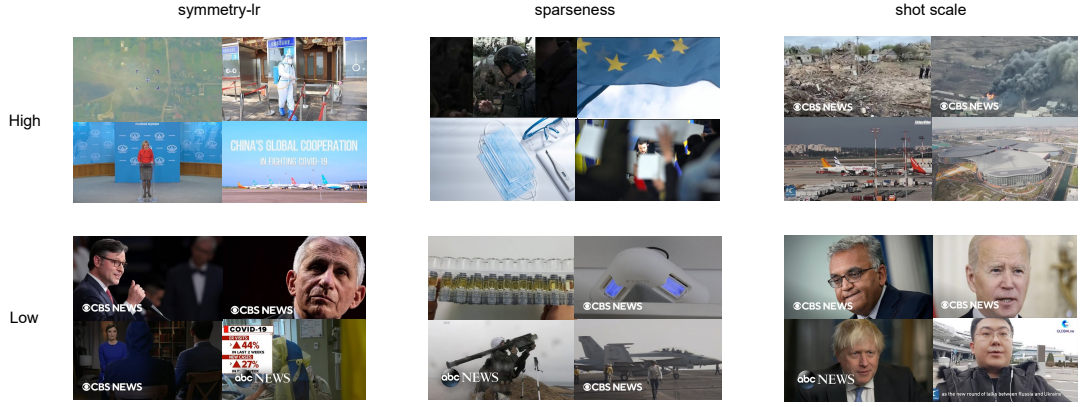}
\caption{Examples of top- and bottom-ranked thumbnails for three of the CNN-based features.}
\label{fig:qualitative_check}
\end{center}
\end{figure*}

\subsubsection{Symmetry.}

\textit{Symmetry-LR, Symmetry-UD}.
Pre-trained Convolutional Neural Network (CNN) models, such as AlexNet, 
generally learn low-level features such as color and texture in their initial convolutional layers.
These activation layers can be used to produce a symmetry metric that is closely aligned with human perception.  
The computation simply compares their values against their flipped versions:
a horizontal flip for 
\textit{lr} (left-right) symmetry (see Figure~\ref{fig:qualitative_check}), and a vertical flip for \textit{ud} (up-down) symmetry.

\subsubsection{CNN.}

\textit{Sparseness, Variability}.
The activation layers of a CNN, such as AlexNet, can also be used to calculate higher level human percepts.

\textit{Sparseness} can be defined by computing
the variances in a CNN's initial response maps. 
Images with high sparseness often correspond to images with homogeneous patches, such as images with only a few objects; see Figure~\ref{fig:qualitative_check}.

\textit{Variability} can be measured as the variance over all response maps, and captures something like the inverse of \textit{self-similarity}, such as with vegetation imagery. 

\subsubsection{Other.}
\textit{Shot Scale, Bag-of-objects}.
we complemented aesthetic features with two additional features reflecting photographic composition.

\textit{Shot scale.}  From the CineScale project,
we used a trained CNN to predict the shot scale of an image as close, medium, or long; see Figure~\ref{fig:qualitative_check}.

\textit{Bag-of-objects.}.
Using the CNN YOLOv5, 
we represented each image as an object histogram, where each bin represents
a labeled object, and the value corresponds to the frequency of that object in the image,
analogous to a bag-of-words model in NLP. 

\section{Validations}

Before further processing, we explored the ability of a feature to create consistent clusters of thumbnails.  


\subsection{Indoor/Outdoor.}
Using the 
PlacesCNN model~\cite{zhou2014learning}, which can label each image as indoor/outdoor, we measured the entropy of each theme using the Gini index of statistical dispersion.

After cleaning the dataset, the Ukraine news story had 599 outdoor images and 575 indoor images with a Gini index of 0.50, a nearly perfect balance. 
Whereas, the COVID-19 news story had 931 indoor and 239 outdoor image with a Gini index of 0.33, mostly due to a strong predominance of medical press conferences. 

For COVID-19, the Gini indices of two out of the three individual themes in Table~\ref{tab:tagging} were less than the aggregate baseline, suggesting that our image clusters had produced groups that were more consistent in setting. This was especially true for the
third ``equipment'' theme being quite pure (0.16) compared to the baseline (0.33).
For Ukraine, the Gini indices also showed a decrease in entropy for all four individual themes in Table~\ref{tab:tagging},
particularly for the ``rubble'' theme (0.26), in which outdoor scenes dominated.

We considered these clustering results to be consistent.

\subsection{CNN features.}
We qualitatively evaluated the extraction of CNN features by visualizing images with the highest/lowest values for certain features. 

Visual examination verified that
images with low left-right symmetry tended to have a dominant subject
framed at the edges of the screen; 
sparse images tended to be of low saturation; and
images 
identified as close-up shots were usually portraits of people. 

\section{Results}

\subsection{Viewership behavior.}
\textit{Do U.S. and Chinese viewers behave differently on YouTube?}
We compared the distribution of viewership statistics (views, likes, comments) of each theme on each channel.

\subsubsection{Views.}
The log-log plots of views count vs. view rank strongly suggested a power law distribution.
All four channels follow roughly the same behavior for the Ukraine event.
For the COVID-19 news event, however, the Chinese channels show a noticeably steeper slope, suggesting a kind of Matthew effect, or possibly a media ``contagion'' effect due to the more immediate personal impact of this event there.

\subsubsection{Likes.}

\begin{figure}[ht]
\begin{center}
\includegraphics[height=4.1cm]{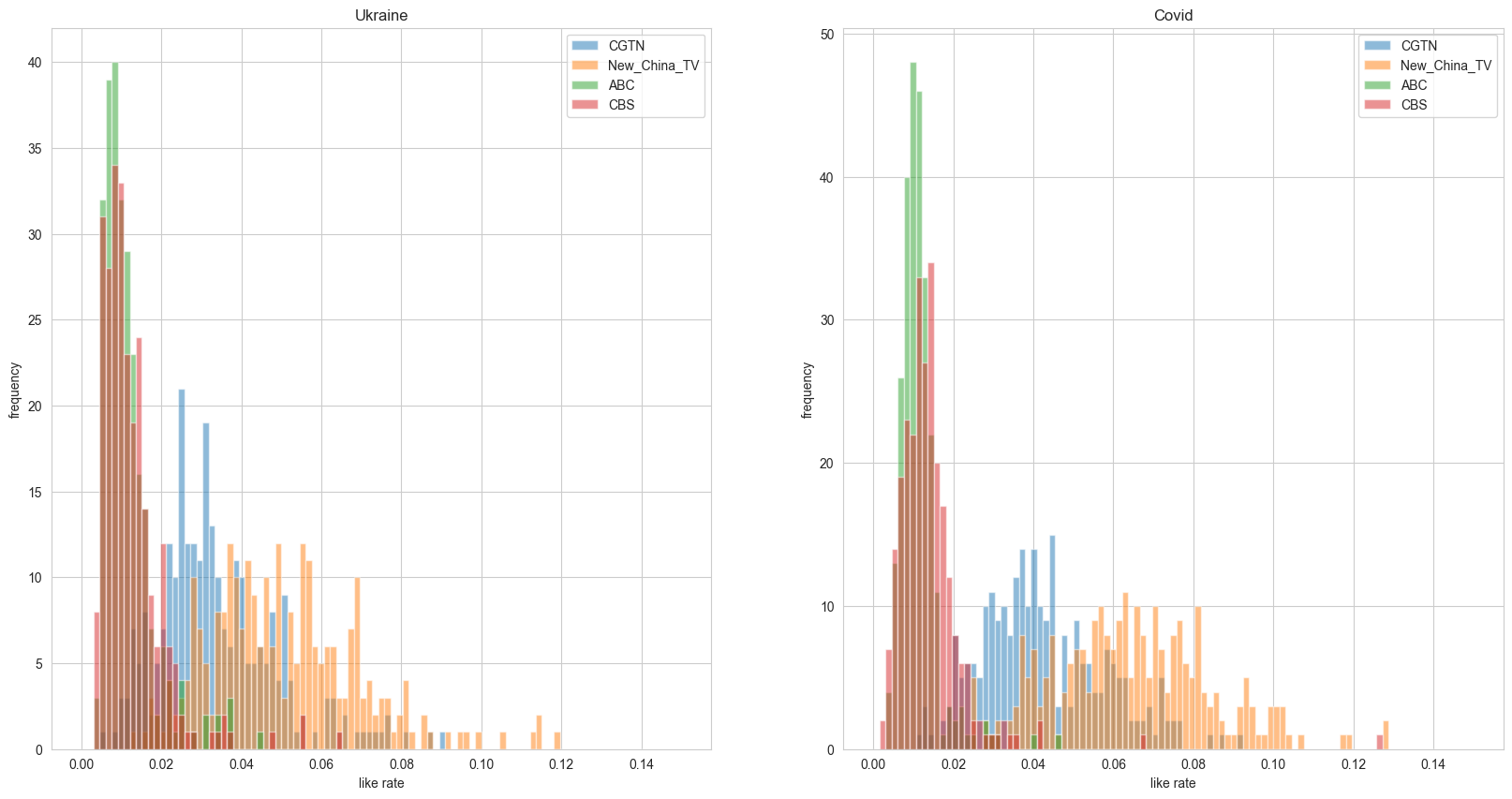}
\caption{Like-rate distribution across events and channels.}
\label{fig:likes_dist}
\end{center}
\end{figure}

We did not find that likes adhered to a power law distribution.
However, we did find that like-{\it rates} (likes divided by views) differed significantly between U.S. vs. Chinese channels; 
see Figure~\ref{fig:likes_dist}. 
U.S. channels generally
have a lower like-rate, even though they
have significantly more views overall.

\subsubsection{Comments.}
We found no detectable differences in the rate at which viewers commented on a video.

\subsection{Theme Preferences.}
\textit{Do U.S. and Chinese channels cover different news themes?}

\subsubsection{Ukraine.}

As shown previously in Figure~\ref{fig:cultural_distributions},
U.S. thumbnails have a noticeably stronger focus on the two militaristic visual themes, whereas
Chinese thumbnails focused more on the two more political visual themes.

This difference may reflect the intentions of the different media outlets, or the different audience preferences, or both.
Chinese channels, due to political affiliation with Russia, 
might be more hesitant to cover the militaristic point of view,
whereas sources based in the U.S., which have openly declared their support for Ukraine, may be more focused on bringing awareness to the human cost of the war.
Alternatively, 
U.S. audiences
may be more drawn to click on thumbnails that involve violence,
and so these channels might cater to this preference to increase their metrics. 

\subsubsection{COVID-19.}
As shown previously in Figure~\ref{fig:cultural_distributions},
both cultures have a large emphasis on the medical context of COVID-19, with the ``mask, equipment, syringe'' visual theme being
dominant.

\subsubsection{Themes over Time.}

\begin{figure}[ht]
\begin{center}
\includegraphics[height=3.9cm]{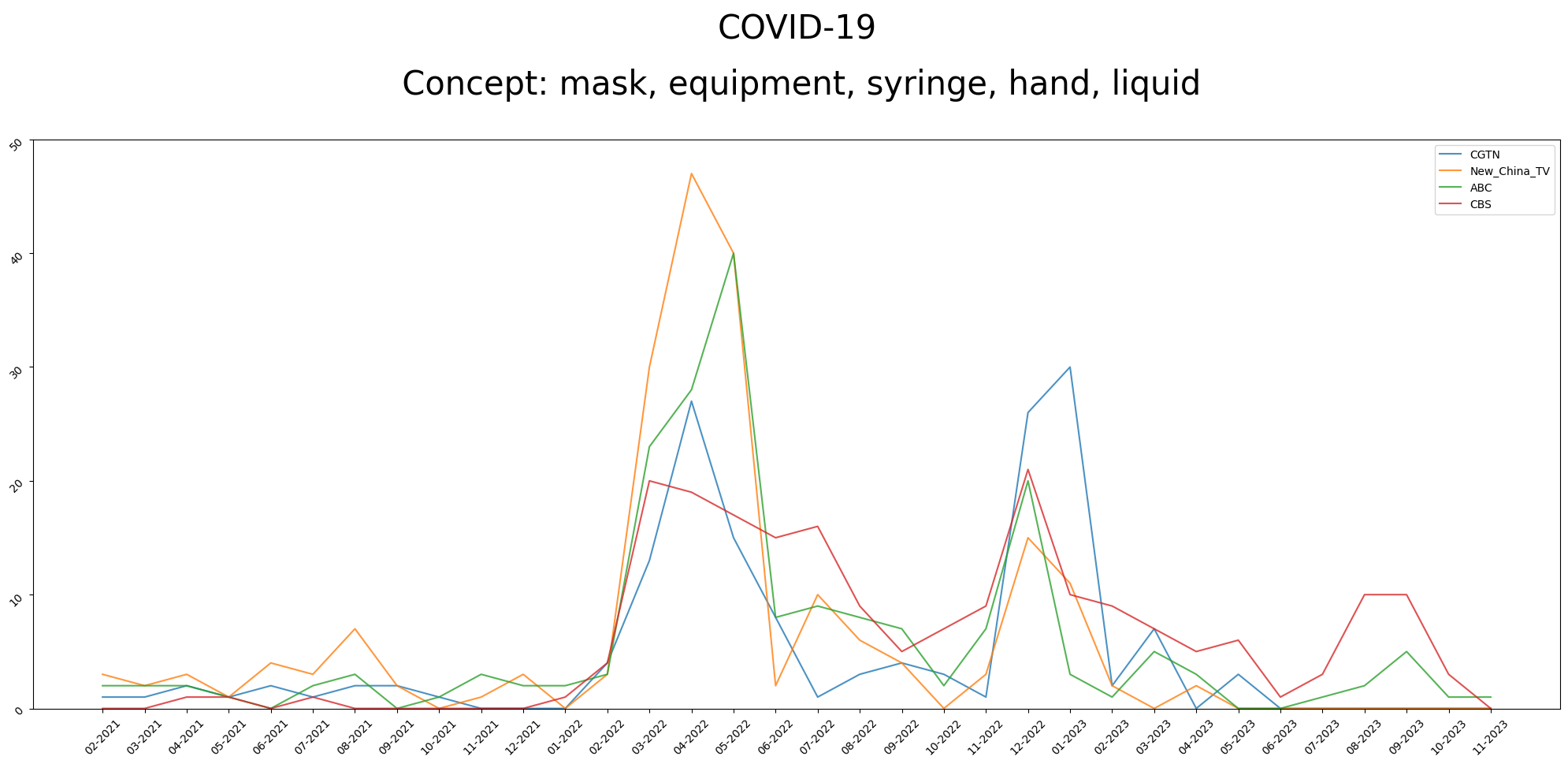}
\caption{Distribution of a COVID theme over time.}
\label{fig:covid_thumbnails_over_time}
\end{center}
\end{figure}

For both events, video uploads peak at times of crisis, and sometimes in a culture-specific way.
For example, Figure~\ref{fig:covid_thumbnails_over_time} traces the COVID ``mask'' visual theme. 
The first and dominant peak occurs at the emergence of the Omicron variant (April, 2022), 
the second at the rescinding of the ``Zero-COVID'' policy in China (December, 2022) after which videos from the Chinese sources drop to zero,
and the third at the Fall 2023 resurgence (September, 2023) reported only in the U.S.

\begin{figure*}[ht]
\begin{center}
\includegraphics[height=3.5cm]{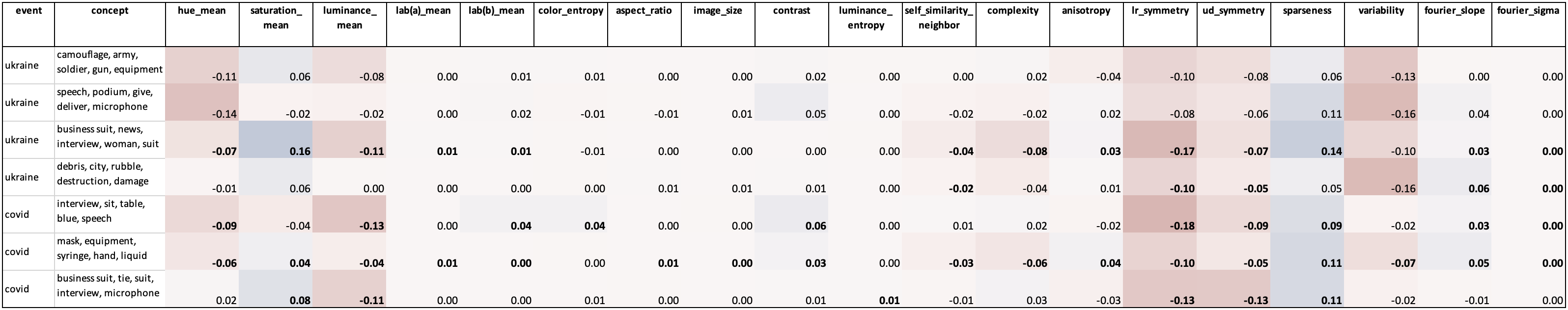}
\caption{Preferences for image aesthetics.  Blue: U.S. is greater.  Red: Chinese is greater. Bold: statistically significant at $p<.05$.}
\label{fig:cultural_aesthetics_2}
\end{center}
\end{figure*}

\subsection{Aesthetic Preferences.}
\textit{Do U.S. vs. Chinese thumbnails differ in their aesthetics?}

\subsubsection{Feature Correlations.}
We derived a feature correlation matrix between each image feature across the entire dataset, and note the following.



\textit{Color} As expected, the HSV and the L*a*b* spaces, which both attempt to match human visual perception, share several strong correlations, particularly with hue.

\textit{HOG}
The features of self-similarity and anisotropy are strongly negatively correlated.
An image with low anisotropy (i.e., one with a lower standard deviation of edge intensities across orientations, due to many parallels within a building or to featureless regions within the sky) 
generates HOG features that
are more likely to be similar across adjacent patches.

\textit{Dimension vs. Fourier}
An inverse relationship between image size and Fourier sigma was unexpected, and is likely an artifact of pre-processing.
Because we had to remove user-added black bars of various sizes, thumbnails
now have different dimensions,
which then affects their spatial frequencies.

\subsubsection{Aesthetic Comparisons, Scalar.}


To examine cultural preferences, we display in Figure~\ref{fig:cultural_aesthetics_2}
the results from a two-sample (U.S. vs. Chinese) t-test for the 19 scalar visual features, for each of the seven visual themes given in Table~\ref{tab:tagging}.

The value in each cell gives the normalized difference (``difference over sum'') between the U.S. and Chinese average feature values for that cell.
Positive values indicate that U.S. thumbnails exhibit the larger average value for the feature, and their cell is shaded blue; negative values in red cells indicate the reverse.
Values are made bold when
the difference between U.S. and Chinese thumbnail distributions is statistically significant ($p < 0.05$).

We note that some features show a cultural preference that generalizes across visual themes,
irrespective of the image content.
These show up in Figure~\ref{fig:cultural_aesthetics_2} as (nearly) uniformly colored columns.

\begin{figure}[ht]
\begin{center}
\includegraphics[height=4.5cm]{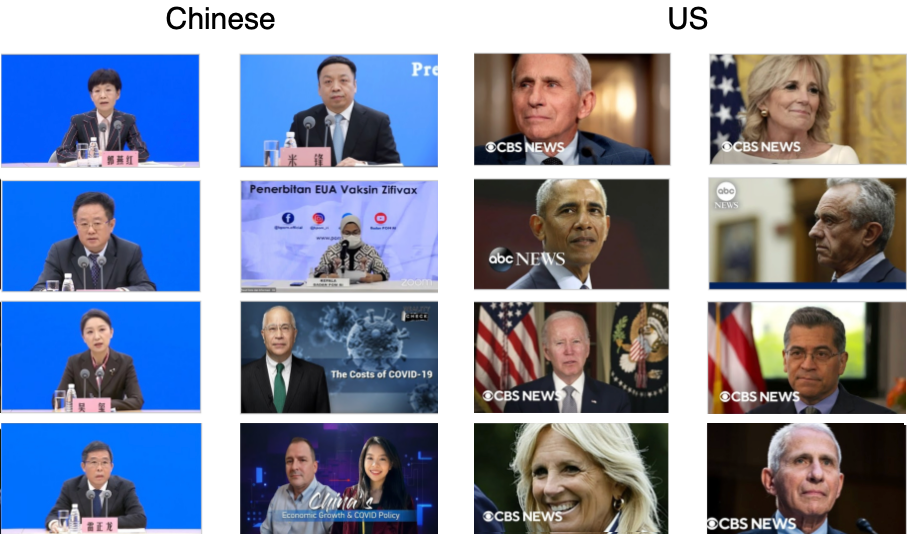}
\caption{Chinese vs. U.S. press conferences; note blue.}
\label{fig:press_conferences}
\end{center}
\end{figure}

\textit{Color: Hue.}
Chinese thumbnails tend to have greater hue values (i.e., they are ``more colorful'') than U.S. thumbnails, 
especially 
for thumbnails depicting interviews or press conferences, which favor featureless but vibrant blue backdrops;
see Figure~\ref{fig:press_conferences}.
U.S. thumbnails for the same visual theme often instead favor
more natural backdrops
such as a room or the White House.




\textit{Color: Saturation.}
Although U.S. images are less ``colorful'',
their colors tend to be more saturated.
This is possibly the result of an editing process, to give the impression that they had been photographed and edited separately,
whereas Chinese thumbnails may strive to be more candid, as if they were taken directly from a frame in the video.

\textit{Lightness: Luminance.}
In general, Chinese thumbnails also tend to be brighter, and appear 
less formal overall, occasionally even overlaying bright animated components on top of background images.

\textit{Fourier: Fourier Slope.}
This metric is greater in U.S. thumbnails, especially in ones relating to ``rubble'' and ``debris'' in the Ukraine news story.
The increased detail may be a result of higher-grade cameras in the field,
further supporting an observation that U.S. thumbnails are more professional,whereas Chinese ones are less formal. 

\textit{Symmetry: -LR, -UD.}
Chinese thumbnails are, on average, more symmetric in both the horizontal and vertical directions, 
across all events, and statistically significantly so
in five of the seven visual themes.
We note
that U.S. thumbnails instead adhere more closely to the photographic ``rule of thirds'' where the foreground
figure is framed more to the left or right.
Chinese thumbnails appear
to come from the actual video footage itself,
which are more candid in their approach, and favor
long balanced landscape shots.


\textit{CNN: Sparseness, Variability.}
U.S. thumbnails are more sparse, across all seven visual themes.
We note that many U.S. thumbnails are portraits taken with cameras set at a low aperture, causing the (usually dark) background to be blurred; blurred regions have low variance. 
This may also explain why Chinese thumbnails
score higher in the variability metric.

\subsubsection{Aesthetic Comparisons, Discrete.} Examples of the two ``Other'' features follow.

\begin{figure}[ht]
\begin{center}
\includegraphics[height=5cm]{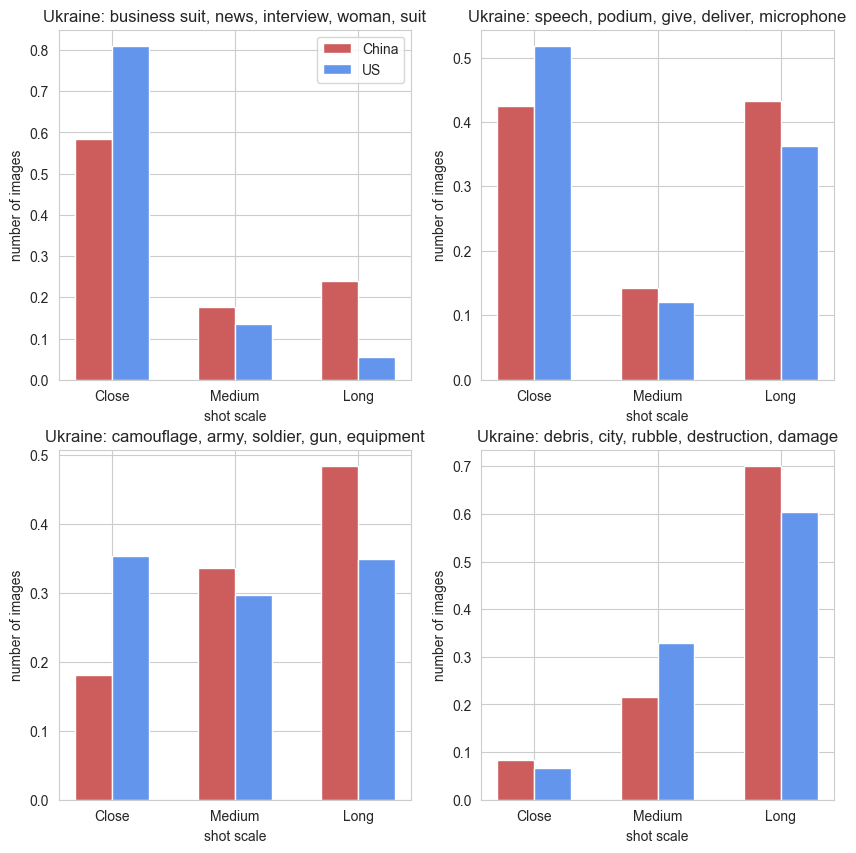}
\caption{Shot scale distributions for the Ukraine themes.}
\label{fig:ukraine_shot_scale}
\end{center}
\end{figure}

\textit{Other: Shot Scale.}
U.S. thumbnails have a stronger preference for close-up shots, portraits in most cases,
whereas Chinese thumbnails favor long shots. 
See Figure~\ref{fig:ukraine_shot_scale} for their comparative use in the four visual
themes in the Ukraine news story.
This may suggest that U.S. channels place more importance on specific individuals making news.

\begin{figure}[ht]
\begin{center}
\includegraphics[height=4cm]{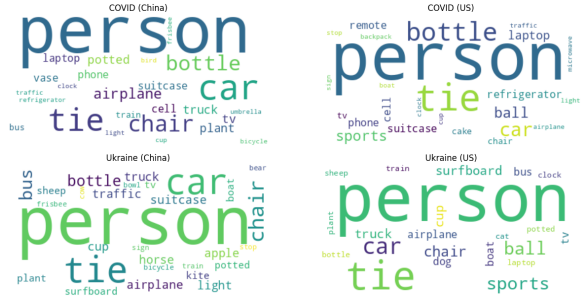}
\caption{Objects found by YOLOv5, event vs. culture.}
\label{fig:objects}
\end{center}
\end{figure}

\textit{Other: Bag of Objects.}
A comparison of the objects that appear most often in thumbnails can be displayed with a word cloud, as in Figure~\ref{fig:objects}, for events vs. cultures. 

All four are dominated by common human-associated objects, such as a person, tie, chair, laptop, TV, bottle, cup, car, or airplane.
Although we found a few event-related differences (refrigerator for COVID, boat and sheep for Ukraine), we found few cultural-related ones (bus for China, sports and ball for U.S.), and did find some errors (surfboard in Ukraine).
This suggests that YoloV5 is not fine-grained enough to detect many culturally-specific objects.

\subsection{Aesthetics vs. Performance.}
\textit{Do U.S. vs. Chinese aesthetics affect usage behavior?}


For each behavioral metric, for each culture, for each for each event, for each visual theme (42 cases in all), we explored how well each of the 19 scalar features was correlated to it (798 correlations in all).
For example, we looked at whether the number of views of Chinese videos about COVID that were tagged as``interview'' were correlated with the strength of the symmetry-ud feature.

For consistency between features, we computed the Spearman rank correlation, over all the videos satisfying the culture, event, and theme.
Then we looked for correlations that were statistically significant.  We found a few.

For example, we found that the above example was in fact positively correlated---and was one of the strongest Chinese correlations of all.
The example could be interpreted as indicating that views increased when there was an isolated talking head shot against a solid featureless background, as in  Figure~\ref{fig:press_conferences} left side, most likely in response to a Chinese crisis.

We also noted that likes increased for U.S. videos about Ukraine that were tagged as ``speech'' if they were dark, sparse, and variable, which again corresponds with the examples in Figure~\ref{fig:press_conferences}, right side, except that these viewers were very responsive. 

\section{Discussion}

Our approach demonstrates a content-controlled analysis of the differences in aesthetic visual features between news thumbnails emerging from sources of varying cultural affinity. 

In summary, we find that U.S. news sources prefer to use professional photographs, whereas Chinese media tend to opt for less formal, more candid images 
taken from the videos themselves.
U.S. thumbnails favor more cinematic close-up portrait shots that strongly employ photographic and editing techniques such as the rule-of-thirds and low apertures,
with deeper color saturation and darker backgrounds. 
In contrast, Chinese thumbnails are often longer shots that are less detailed, 
more colorful, and brighter, suggesting 
a more natural-looking image, which are sometimes
superimposed with animated content.


Besides image aesthetics, these channels also show some cultural differences in the visual themes they depict in thumbnails. 
For the Ukraine conflict, U.S. sources more strongly emphasize the militaristic narrative, favoring scenes with soldiers and destruction, while Chinese sources more frequently cover the geopolitical themes of the conflict, using thumbnails from conference rooms and speeches.


\section{Application to Disinformation}


Previous investigation has shown that multimodal disinformation that uses a text-plus-visual format is perceived as more credible than pure text~\cite{hameleers2020picture}. 
Visual disinformation can involve visual content that is deliberately manipulated or generated, or 
real but
used out of context, or
real but used in tandem with false 
claims~\cite{cao2020exploring}.
This paper focuses on the first two.

\textit{Methodology.} Our observations offer a way to classify and identify the framing strategies that are used in the visual news domain.
Clustering images into visual \textit{themes} can help
illuminate what is being inappropriately stressed or omitted. 
Analyzing visual \textit{features} can help identify attempts to inappropriately manipulate feelings and perceptions.

\textit{Cultural Baselines.}
We have also demonstrated that detecting visual disinformation
would be complicated, since different cultures adhere to different preferences and baselines.
For example, the U.S. audience appears to approach news partly as
a form of interactive engagement,
whereas Chinese viewers appear to approach 
news as more one-way and objective.





To better establish these methods and baselines, further research is necessary.  Some technical issues remain (e.g., removing image watermarks, testing alternatives to Cluster).  More importantly, datasets of larger scale (more videos, events, cultures), particularly from media such as entertainment that encourage user involvement, would help formalize what themes and what features are most critically involved with disinformation.

\clearpage
\bibliography{references}

\clearpage

\section*{Ethics Checklist}

\begin{enumerate}


\item For most authors...
\begin{enumerate}
    \item  Would answering this research question advance science without violating social contracts, such as violating privacy norms, perpetuating unfair profiling, exacerbating the socio-economic divide, or implying disrespect to societies or cultures?
    \answerYes{Yes}
  \item Do your main claims in the abstract and introduction accurately reflect the paper's contributions and scope?
    \answerYes{Yes}
   \item Do you clarify how the proposed methodological approach is appropriate for the claims made? 
    \answerYes{Yes}   
   \item Do you clarify what are possible artifacts in the data used, given population-specific distributions?
    \answerYes{Yes}   
  \item Did you describe the limitations of your work?
    \answerYes{Yes} 
  \item Did you discuss any potential negative societal impacts of your work?
    \answerNA{No. The work is purely observational.}
  \item Did you discuss any potential misuse of your work?
    \answerNA{No. No system was designed or created.}
  \item Did you describe steps taken to prevent or mitigate potential negative outcomes of the research, such as data and model documentation, data anonymization, responsible release, access control, and the reproducibility of findings?
    \answerNo{No. Data is publicly available from news channels and has not identified individuals without their consent.}
  \item Have you read the ethics review guidelines and ensured that your paper conforms to them?
    \answerYes{Yes}
\end{enumerate}

\item Additionally, if your study involves hypotheses testing...
\begin{enumerate}
  \item Did you clearly state the assumptions underlying all theoretical results?
    \answerNA{No theoretical results.}
  \item Have you provided justifications for all theoretical results?
    \answerNA{No theoretical results.}
  \item Did you discuss competing hypotheses or theories that might challenge or complement your theoretical results?
    \answerNA{No theoretical results.}
  \item Have you considered alternative mechanisms or explanations that might account for the same outcomes observed in your study?
    \answerYes{Yes}
  \item Did you address potential biases or limitations in your theoretical framework?
    \answerYes{Yes. Dataset is limited in size.}
  \item Have you related your theoretical results to the existing literature in social science?
    \answerNA{No theoretical results.}
  \item Did you discuss the implications of your theoretical results for policy, practice, or further research in the social science domain?
    \answerNA{No theoretical results.}
\end{enumerate}

\item Additionally, if you are including theoretical proofs...
\begin{enumerate}
  \item Did you state the full set of assumptions of all theoretical results?
    \answerNA{No theoretical proofs.}
  \item Did you include complete proofs of all theoretical results?
    \answerNA{No theoretical proofs.}
\end{enumerate}

\item Additionally, if you ran machine learning experiments...
\begin{enumerate}
  \item Did you include the code, data, and instructions needed to reproduce the main experimental results (either in the supplemental material or as a URL)?
    \answerNo{No. No substantive code was created, and the off-the-shelf CNN is cited.  Data is publicly available via YouTube API.}
  \item Did you specify all the training details (e.g., data splits, hyperparameters, how they were chosen)?
    \answerYes{Yes}
  \item Did you report error bars (e.g., with respect to the random seed after running experiments multiple times)?
    \answerNA{No. Experiments were statistical and run once. }
  \item Did you include the total amount of compute and the type of resources used (e.g., type of GPUs, internal cluster, or cloud provider)?
    \answerNA{No. No training of any system was involved.}
     \item Do you justify how the proposed evaluation is sufficient and appropriate to the claims made? 
    \answerYes{Yes.}
     \item Do you discuss what is ``the cost`` of misclassification and fault (in)tolerance?
    \answerNA{No. The paper notes preferences and no hard decisions.}
  
\end{enumerate}

\item Additionally, if you are using existing assets (e.g., code, data, models) or curating/releasing new assets, \textbf{without compromising anonymity}...
\begin{enumerate}
  \item If your work uses existing assets, did you cite the creators?
    \answerYes{Yes}
  \item Did you mention the license of the assets?
    \answerNA{No. The system (Cluster) and data source (YouTube) are common and well-known.}
  \item Did you include any new assets in the supplemental material or as a URL?
    \answerNA{No new assets.}
  \item Did you discuss whether and how consent was obtained from people whose data you're using/curating?
    \answerNA{No consent was required.}
  \item Did you discuss whether the data you are using/curating contains personally identifiable information or offensive content?
    \answerNA{No. Data was standard YouTube download off well-established news channels.}
  \item If you are curating or releasing new datasets, did you discuss how you intend to make your datasets FAIR (see \citet{fair})?
    \answerNA{No new datasets.}
  \item If you are curating or releasing new datasets, did you create a Datasheet for the Dataset (see \citet{gebru2021datasheets})? 
    \answerNA{No new datasets.}
\end{enumerate}

\item Additionally, if you used crowdsourcing or conducted research with human subjects, \textbf{without compromising anonymity}...
\begin{enumerate}
  \item Did you include the full text of instructions given to participants and screenshots?
    \answerNA{No human subjects.}
  \item Did you describe any potential participant risks, with mentions of Institutional Review Board (IRB) approvals?
    \answerNA{No human subjects.}
  \item Did you include the estimated hourly wage paid to participants and the total amount spent on participant compensation?
    \answerNA{No human subjects.}
  \item Did you discuss how data is stored, shared, and deidentified?
   \answerNA{No human subjects.}
\end{enumerate}

\end{enumerate}

\end{document}